\documentclass[a4paper,11pt]{article}
\usepackage[utf8]{inputenc}
\usepackage{graphicx}
\usepackage{amsmath}
\pdfoutput=1
\usepackage{slashed,mathtools}
\usepackage{jheppub}
\usepackage{subfig}
\usepackage{multirow}
\usepackage{tikz}
\usepackage{enumitem}
\usepackage{tablefootnote}
\usepackage{cases}
\allowdisplaybreaks

\title{Exchange driven freeze out of dark matter}

\author[a]{Tarak Nath Maity,}
\author[a,b]{Tirtha Sankar Ray}
\affiliation[a]{Department of Physics, Indian Institute of Technology Kharagpur, Kharagpur 721302, India}
\affiliation[b]{Centre for Theoretical Studies, Indian Institute of Technology Kharagpur, Kharagpur 721302, India}
\emailAdd{tarak.maity.physics@gmail.com}
\emailAdd{tirthasankar.ray@gmail.com}

\abstract{We introduce a novel mechanism where processes that preserve the  number density  of the  dark sector  set the relic density of a thermal particulate dark matter. In a relatively degenerate multipartite dark sector if there is a considerable time lapse between the freeze out of various species then a process like exchange between dark sector constituents can play the pivotal role of driving freeze out and setting dark matter relic density. As a proof of principle we  present simple scalar models with viable  dark matter  in the GeV scale to demonstrate this phenomenon.}


\begin{document}

\maketitle
\flushbottom

\section{Introduction}
\label{sec:intro}
The overwhelming evidence from various astrophysical and cosmological observations establish that around $25\%$ of the energy budget of the universe is made of dark matter (DM) \cite{Profumo:2017hqp, Doroshkevich:1984gw}. The assumption that DM is a particulate thermal relic is one of the leading candidate.  In this framework DM remains in thermal equilibrium at very early stage of the thermal history of the Universe, however, due to the combined effect of depletion in its number density through annihilation and the expansion of the Universe, it gets frozen out resulting in the DM relic density observable today.

\begin{figure}[t]
	\begin{tikzpicture}[line width=1.5 pt, scale=2]
		\subfloat[\label{sf:ann}]{\draw (145:1) -- (145:.3cm);
			\node at (145:1.15) {$\phi_1$};
		\draw (215:1) -- (215:.3cm);
			\node at (215:1.15) {$\phi_1$};
		\draw (35:1) -- (35:.3cm);
			\node at (35:1.2) {SM};
		\draw (-35:1) -- (-35:.3cm);
			\node at (-35:1.17) {SM};
		\draw[fill=black] (0,0) circle (.3cm);
		\draw[fill=white] (0,0) circle (.29cm);
		\begin{scope}
	    	\clip (0,0) circle (.3cm);
	    	\foreach \x in {-.9,-.8,...,.3}
				\draw[line width=1 pt] (\x,-.3) -- (\x+.6,.3);
	  	\end{scope}
	  	\node at (0,-1) {(a)};}
	  	
	    \subfloat[\label{sf:exchange}]{\hspace{5cm}
		\draw (145:1) -- (145:.3cm);
		\node at (145:1.15) {$\phi_1$};
		\draw (215:1) -- (215:.3cm);
		\node at (215:1.15) {$\phi_1$};
		\draw (35:1) -- (35:.3cm);
		\node at (35:1.15) {$\phi_2$};
		\draw (-35:1) -- (-35:.3cm);
		\node at (-35:1.17) {$\phi_2$};
		\draw[fill=black] (0,0) circle (.3cm);
		\draw[fill=white] (0,0) circle (.29cm);
		\begin{scope}
		\clip (0,0) circle (.3cm);
		\foreach \x in {-.9,-.8,...,.3}
		\draw[line width=1 pt] (\x,-.3) -- (\x+.6,.3);
		\end{scope}	
	  	\node at (0,-1) {(b)};}
		\subfloat[\label{sf:ann2}]{
			    \hspace{5cm}
			    \draw (145:1) -- (145:.3cm);
			    \node at (145:1.15) {$\phi_2$};
			    \draw (215:1) -- (215:.3cm);
			    \node at (215:1.15) {$\phi_2$};
			    \draw (35:1) -- (35:.3cm);
			    \node at (35:1.2) {SM};
			    \draw (-35:1) -- (-35:.3cm);
			    \node at (-35:1.17) {SM};
			    \draw[fill=black] (0,0) circle (.3cm);
			    \draw[fill=white] (0,0) circle (.29cm);
			    \begin{scope}
			    \clip (0,0) circle (.3cm);
			    \foreach \x in {-.9,-.8,...,.3}
			    \draw[line width=1 pt] (\x,-.3) -- (\x+.6,.3);
			    \end{scope}	 
	  	\node at (0,-1) {(c)};}	
	  \end{tikzpicture}
\begin{center}
\begin{tikzpicture} 
\draw[line width=0.5mm,->] (-5,0) -- node[midway,fill=white] {Freeze out Time} (5,0); 
\end{tikzpicture}
\end{center}
\caption{The main processes that are operative at the time DM freeze out in a multipartite dark sector. The arrow below indicates the decoupling time ordering of the processes that may lead to  exchange driven freeze out of DM. While for reverse freeze-out time ordering between (b) and (c) the annihilation of $\phi_2$ will drive the freeze out of $\phi_1$.}
\label{fig:block-diagram}
\end{figure}
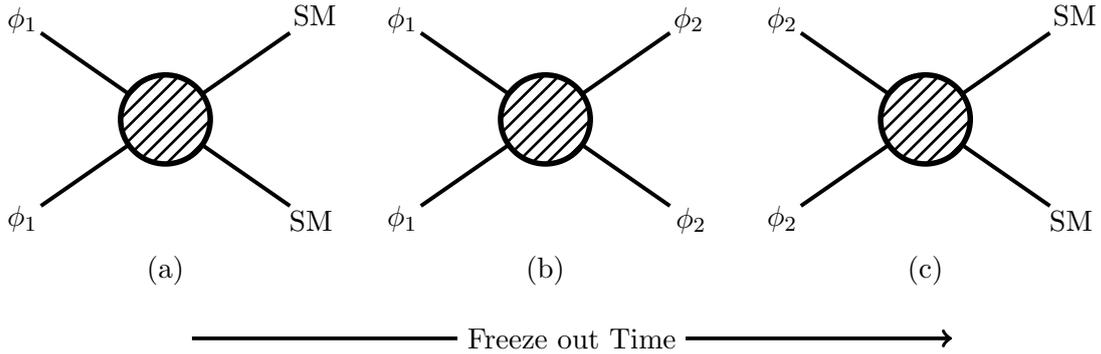

The description of thermal freeze out of DM is usually driven by $2 \to 2$ number changing processes, where a pair of stable DM candidate annihilates to the Standard Model (SM) states \cite{Lee:1977ua, Griest:1990kh, Edsjo:1997bg, Scherrer:1985zt, Srednicki:1988ce, Gondolo:1990dk, Steigman:2012nb}. It so happens that a particle having weak scale mass and coupling could produce the observed relic density \cite{Aghanim:2018eyx} and this outcome is often referred to as WIMP (Weakly Interacting Massive Particles) miracle.  This framework has received considerable attention within a large class of particle physics models \cite{Jungman:1995df, Bertone:2004pz, Arcadi:2017kky} and has been explored widely. However, the framework of WIMP is progressively in tension by the non-observations of it in different dark matter experiments.  Subsequently there have been new proposals like $ 3 \to 2 $ annihilation \cite{Hochberg:2014dra, Dey:2016qgf, Dey:2018yjt}, Secluded DM \cite{Pospelov:2007mp}, Co-decaying DM \cite{Dror:2016rxc}, Co-scattering \cite{Garny:2017rxs, Garny:2019kua, DAgnolo:2017dbv}, Forbidden DM \cite{DAgnolo:2015ujb} etc that addresses some of the shortcomings of the vanilla WIMP framework. Nevertheless, it is not surprising to expect these models to have a multi-particle dark sector \cite{Drozd:2011aa, Bhattacharya:2017fid, Casas:2017jjg, Bhattacharya:2016ysw, Khlopov:1995pa} whose stability is  ensured by imposing one or more discrete symmetry.  Depending on the number of particles and unbroken discrete symmetries in the theory either there will be a single or multi component DM. Assuming that, these dark sector particles couple to SM states in the primordial soup, then, the main underlying process that set the thermal  DM relic density is shown in figure \ref{fig:block-diagram}. To keep the discussion tractable we assume that there are two  particles in the dark sector $\phi_1$  and $\phi_2$ with the former being the major component of the present day DM relic density. Depending on the stabilizing symmetry, $\phi_2$ may either be a minor component of the DM or an associated  unstable state of the dark sector. We will be considering both of these scenarios later.

The chronology of the decoupling time of the processes, shown in figure \ref{fig:block-diagram}, will determine freeze out behaviour of the species under consideration. In the standard paradigm the freeze out of the processes in figure \ref{fig:block-diagram} happens almost simultaneously and the annihilation channels of individual species (figures  \ref{fig:block-diagram}(a) and  \ref{fig:block-diagram}(c) for $\phi_1$ and $\phi_2$ respectively) are primarily responsible for setting their relic density \cite{Griest:1990kh, Edsjo:1997bg, Scherrer:1985zt, Srednicki:1988ce, Gondolo:1990dk, Steigman:2012nb}. It has been already pointed out that, if there is a substantial gap in the freeze out time of these processes  then non-annihilation channels may  regulate the relic density of DM \cite{Garny:2017rxs, Garny:2019kua, DAgnolo:2017dbv}. It is in this context, we propose for the first time a mechanism where exchange processes (shown in figure \ref{fig:block-diagram}(b)) play  the pivotal role in setting DM relic density.

In section \ref{sec:formalism}, we describe basic formalism of exchange driven freeze out mechanism. In section \ref{sec:set-up}, we present simple scalar DM model where the framework has been realized, before concluding in section \ref{sec:con}. 

\section{Basic framework}
\label{sec:formalism}
In this section, we will lay out the basis formalism of an exchange driven freeze out mechanism. Within this framework we will assume that  the  annihilation of the major DM constituent $\phi_1,$  shown in figure \ref{fig:block-diagram}(a),  decouples earliest in the thermal history compared to all the other relevant processes. This early freeze out  implies a  suppression in the annihilation cross section, which may be indicative of  the null results in the DM  experiments. The successful thermal freeze out of $\phi_1$ can  still be achieved through the combined effect of the exchange $\phi_1 \phi_1 \to \phi_2 \phi_2$ and subsequent annihilation of the $\phi_2$ to SM state. This is in contrast with  studies of multi-particle dark sector \cite{Finkbeiner:2009mi,Batell:2009vb,Bhattacharya:2016ysw}, where the relic density is primarily set by annihilation processes.   Interestingly reference \cite{Kopp:2016yji} has considered a related but distinct  framework,  where the DM states  annihilate to  relatively heavier unstable  intermediate  states  that subsequently decay to SM.  The endothermic annihilation of the DM to the intermediate state controls its relic density.

The co-moving number density of $\phi_1$ freezes  once either of  both processes turns in-operational. Crucially, the mode of freeze out hinges on the time ordering of decoupling of processes shown in figure \ref{fig:block-diagram}(b) and \ref{fig:block-diagram}(c).  If the interaction rate of exchange process goes below the Hubble expansion rate before the other one, then the exchange process will determine the freeze out. This is an atypical feature of this framework where a  non-number changing process  drives the freeze out. While for the reverse time ordering, annihilation of $\phi_2$ will regulate the freeze out of $\phi_1$. Interestingly, here the  decoupling of the annihilation of one species sets the relic abundance of a different state.

The evolution of the number density ($n$) of the dark sector species can be obtained by solving the  coupled Boltzmann equation for the dark sector states, that can be schematically written as
\begin{eqnarray}
\label{eq:fullBE}
\dot n_i + 3 \mathcal{H} n_i = - \sum_{j} \Bigg\{ \left< \sigma v \right>_{i}^{\rm anh} \left( n_i^2 - \left(n_i^{\rm eq}\right)^2 \right)
+\left< \sigma v \right>_{i \to j}^{\rm ex} \left( n^2_i-\left(n_i^{\rm eq}\right)^2 \Big(\frac{n_j}{n_j^{\rm eq}}\Big)^2  \right)   \Bigg\},
\end{eqnarray}
where $i,~j$ include all the particles in the spectrum, $n^{\rm eq}_k$  represents the equilibrium number density, $\mathcal{H}$ is the Hubble expansion rate. The first term in the RHS of equation \eqref{eq:fullBE} is related to annihilation processes with $\left< \sigma v \right>_{i}^{\rm anh}$ being the thermally averaged annihilation cross section, and the second term is the contribution of exchange process. 

As is usually the case, if the major DM  component $\phi_1$ is the lightest state in the dark sector, then   the $\phi_1 \phi_1 \to \phi_2 \phi_2$ exchange process is endothermic and its thermally averaged cross section has an exponential suppression \cite{DAgnolo:2017dbv}\footnote{Such forbidden channel has been considered earlier in the context of the annihilation of sub-GeV DM \cite{DAgnolo:2015ujb} and in the co-scattering scenario \cite{ DAgnolo:2017dbv}.}.
\begin{equation}
\label{eq:x-section}
\langle \sigma v \rangle_{1 \to 2}^{\rm ex} \approx \Big(\frac{m_2}{m_1}\Big)^3 e^{- 2x \delta}\langle \sigma v \rangle_{2 \to 1},
\end{equation}
where $x=m_1/T$, and $\delta=(m_2-m_1)/m_1$ is a measure of the degree of degeneracy in the dark sector spectrum. 

In the exchange driven freeze out regime we can keep aside all the suppressed cross sections and considering the fact that the $\phi_2$ remains in equilibrium at the time of freeze out of $\phi_1$, equation \eqref{eq:fullBE} reduces to 
\begin{equation}
\dot n_1 + 3 \mathcal{H} n_1 = -  \left< \sigma_{11 \rightarrow 22} v \right>   \left( n^2_1 - \left(n_1^{\rm eq}\right)^2 \right)
\label{eq:exBE}
\end{equation}
Interestingly, equation \eqref{eq:exBE} is quite similar to the Boltzmann equation for the traditional WIMPs. The solution of equation \eqref{eq:exBE} has already been studied extensively in the literature \cite{Kolb:1990vq, Scherrer:1985zt, Srednicki:1988ce, Gondolo:1990dk, Steigman:2012nb}.    

On the other hand, if the annihilation of $\phi_2$ freezes before the exchange process, then the freeze out of $\phi_1$ is determined by the related freeze out of $\phi_2$\footnote{This mechanism is similar to the sterile co-annihilation framework discussed in the context of light dark matter \cite{DAgnolo:2018wcn}}. Therefore, retaining relevant terms, the equation \eqref{eq:fullBE} trims down to following coupled system of differential equations:
\begin{subequations}
\begin{align}
\dot n_1 + 3 \mathcal{H} n_1 = & -  \left< \sigma_{11 \rightarrow 22} v \right>   \left( n^2_1 - \left(n_1^{\rm eq}\right)^2 \left(\frac{n_2}{n_2^{\rm eq}}\right)^2\right) \\
\dot n_2 + 3 \mathcal{H} n_2 =& -  \left< \sigma_{22} v \right>   \left( n^2_2 - \left(n_2^{\rm eq}\right)^2 \right)
\label{eq:NLSP-BE}
\end{align}
\end{subequations}
The coupled Boltzmann equation \eqref{eq:NLSP-BE} can be solved numerically to get the required relic abundance.

An approximate  analytical solution of the coupled Boltzmann equation \eqref{eq:fullBE}  in  the exchange and $\phi_2$ annihilation dominated  regime  may be obtained by comparing the relevant event rate with the Hubble parameter through the following equation:
\begin{eqnarray}
n_{1} \langle \sigma v \rangle^{\rm ex} &=& \mathcal{H}, \quad \rm exchange \nonumber \\
n_{2} \langle \sigma v \rangle^{\rm anh} &=& \mathcal{H}, \quad \phi_2~ \rm annihilation
\end{eqnarray}

This gives an estimate of the freeze out temperature $(x_f)$ of the $\phi_1$ up to  a  normalizing factor $(c)$ in the event rate. 
\begin{numcases}
{x_f \approx } \frac{1}{2\delta}\ln \left(  \frac{M_{\rm pl} \, m_1 \langle \sigma v \rangle_{2 \to 1}^{\rm ex} (1+3 \delta) x_f^{1/2} }{26.14 c \sqrt{g_{\rm eff}(x_f)}  }\right), \quad   ~ \rm exchange \nonumber \\
\frac{1}{\delta+1}\ln \left(  \frac{M_{\rm pl} \, m_1 \langle \sigma v \rangle_{2}^{\rm anh} (1+3\delta/2 )x_f^{1/2} }{26.14 c \sqrt{g_{\rm eff}(x_f)} }\right),\quad   \phi_2~ \rm annihilation 
\label{eq:xf}
\end{numcases}
Note that, in equation \eqref{eq:xf} $g_{\rm eff}$ is the effective number of relativistic degrees of freedom at the time of freeze out.

The relic density can now be computed assuming that the co-moving number density of the $\phi_1$ remains constant since freeze out. The approximate expression for the major component of DM relic density for the two regime can be written as,
%
%
\begin{numcases}
{\Omega h^2 \approx }
\frac{0.27~ \rm pb}{\langle \sigma v \rangle_{2 \to 1}^{\rm ex}} \frac{c \, x_f \, e^{x_f (2\delta-1)} \,(1-3\delta)}{\sqrt{g_{\rm eff}(x_f)} },\quad \rm{ exchange}  \nonumber \\
\frac{0.27~ \rm pb}{\langle \sigma v \rangle_{2}^{\rm anh}} \frac{c \, x_f \, e^{x_f \delta} (1-3 \delta/2)}{\sqrt{g_{\rm eff}(x_f)} \, }, \quad \rm{ \phi_2~ annihilation} 
\label{eq:relic}
\end{numcases}
%
%
Matching with a full numerical solution of equation \eqref{eq:fullBE}  implies $c \sim \mathcal{O}(1).$  With $g_{\rm eff} \sim 100$ from equation \eqref{eq:xf}, the experimentally observed value of relic density could be reproduce for $x_f \sim 20$, implying DM freeze out  in the non-relativistic regime. This signifies that exchange driven  and $\phi_2$  annihilation driven freeze out yield a cold dark matter as preferred from the large scale structure of the Universe.

\section{Minimal setup}
\label{sec:set-up}
To construct the minimal multi-particle dark sector, we augment the SM with two real scalar SM singlet fields $\phi_1$ and $\phi_2$. Depending on the stabilizing symmetry of the dark sector one can obtain a single component or two component DM.  Regions of parameter space of these generic constructions demonstrate  the phenomenon of exchange driven freeze out.

\subsection{Model 1: Multi-component DM}
\label{subsec:model1}
The most economic Lagrangian is obtained when one considers two discrete symmetries $\mathbb{Z}_2$ and $\mathbb{Z}_2^{\prime}$  under which $\phi_1$ and $\phi_2$ are charged, respectively, while the SM remains even under both \cite{Bhattacharya:2016ysw}. The most general renormalizable Lagrangian including the  Higgs portal coupling to the SM can be written as
\begin{figure}[t]
\begin{center}
\includegraphics[scale=0.27]{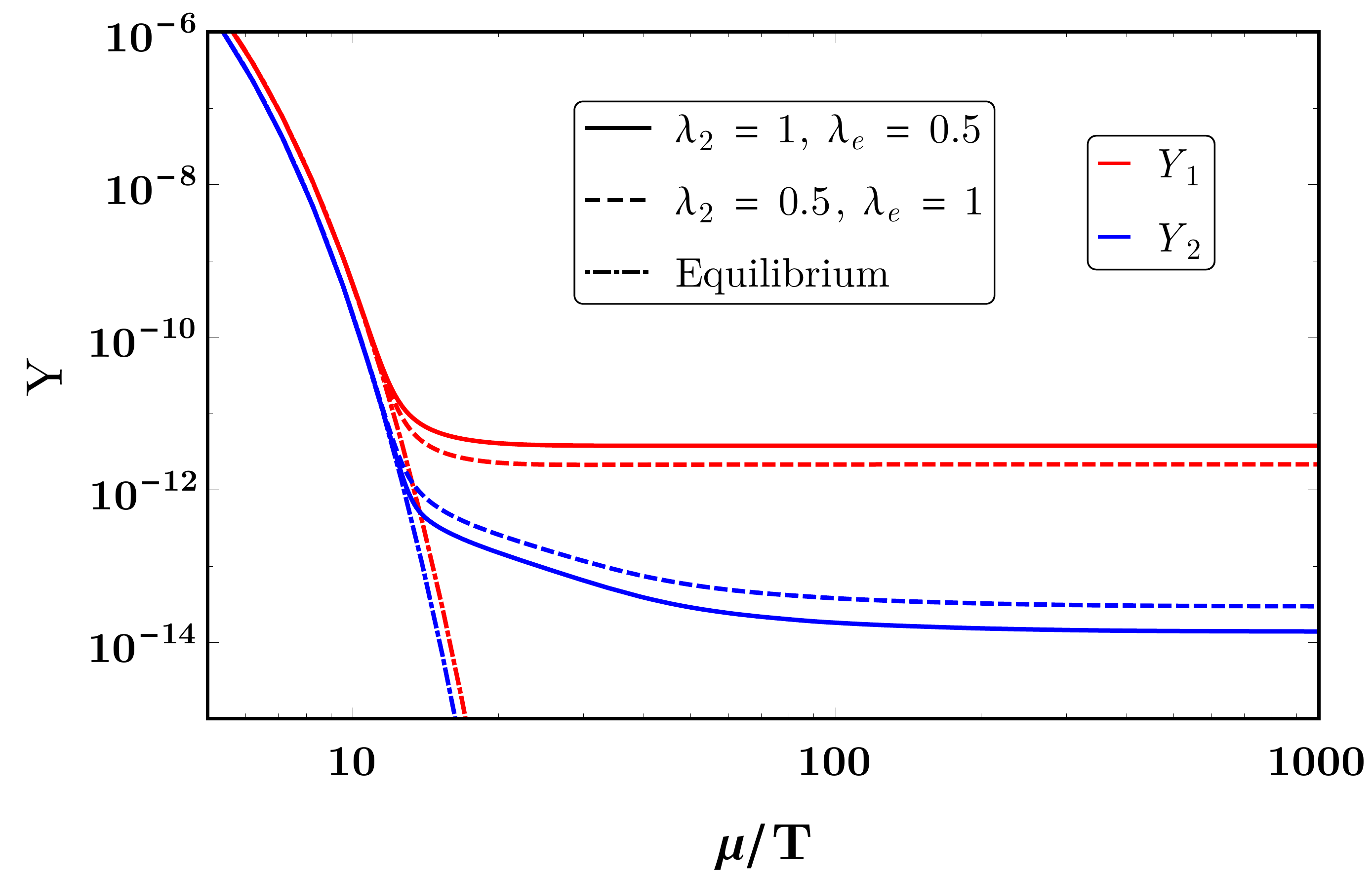}
\caption{Evolution of co-moving number densities with time. The red and blue lines indicate evolution of co-moving number densities of $\phi_1$ and $\phi_2$. For the solid and dashed lines, exchange and $\phi_2$ annihilation processes drive the decoupling of $\phi_1$. Note that $\mu$ is the reduced mass of two dark matter. The chosen values of $m_1$ and $m_2$ are $100$ and $105$ GeV, respectively. We set  $\lambda_1 = 10^{-5} $ for all the contours.}
\label{fig:yt}
\end{center}
\end{figure}
 \begin{eqnarray}
\label{eq:z2z2pL}
-\mathcal{L}_{\rm int} = \frac{\lambda_{1}}{2}\phi_1^2 H^{\dagger}H  +\frac{\lambda_{1_s}}{4!}\phi_1^4+\frac{\lambda_{2_s}}{4!}\phi_2^4 +\frac{\lambda_{2}}{2} \, \phi_2^2   H^{\dagger}H+\frac{\lambda_{e}}{4}\phi_1^2\phi_2^2 ,  
\end{eqnarray} 
where $H$, denotes the SM Higgs doublet. In writing above equation, we have omitted mass term for $\phi_1$ and $\phi_2$. We assume the masses of $\phi_1$ and $\phi_2$ are $m_1$ and $m_2$, respectively. Clearly, both $\phi_1$ and $\phi_2$ will be stable leading to a two component  DM.

\begin{figure*}[t]
\centering
\subfloat[\label{sf:Ex-regime}]{\includegraphics[width=.45\linewidth,height=5cm]{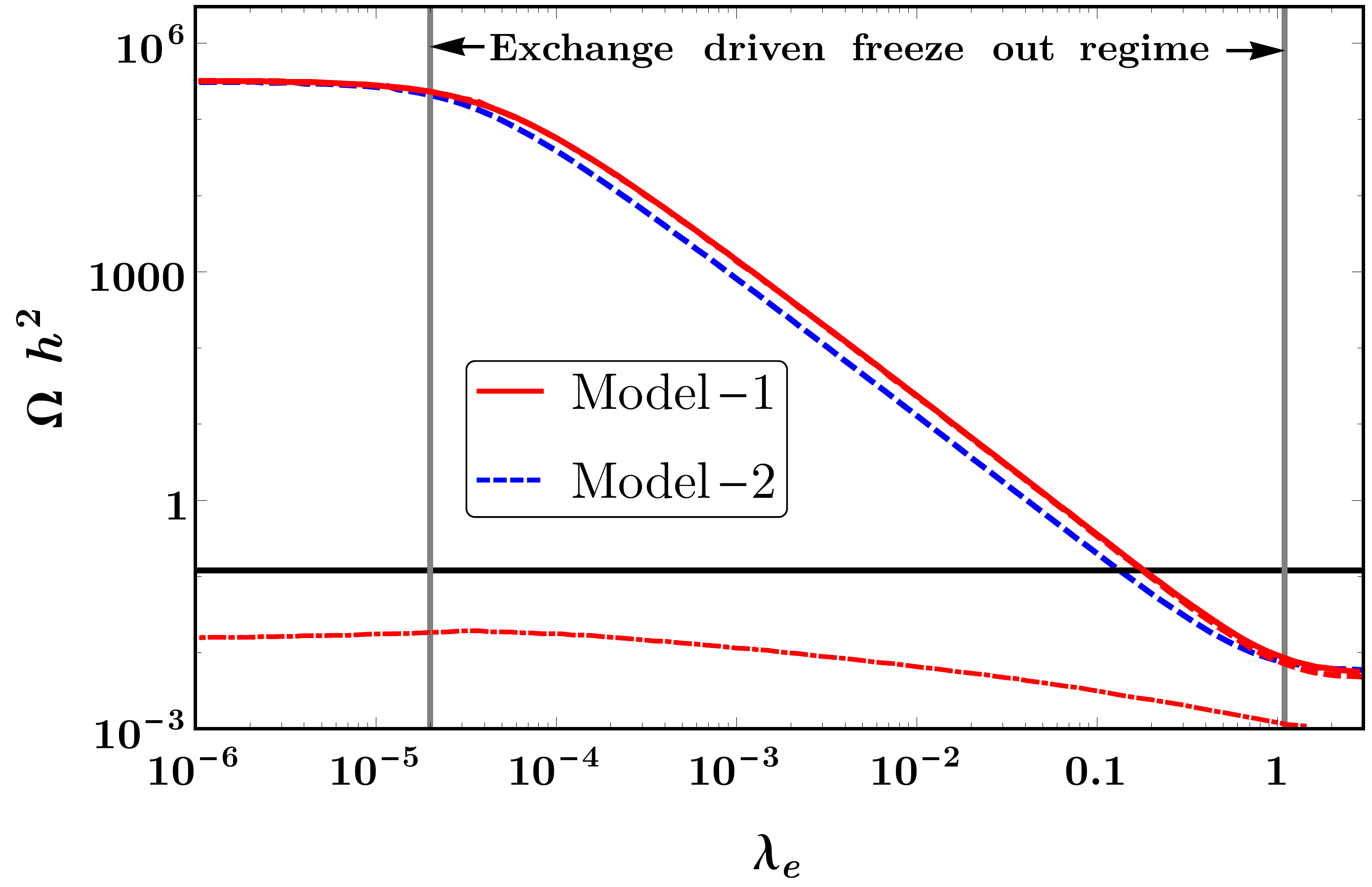}}\quad 
\subfloat[\label{sf:2ndAnn-regime}]{\includegraphics[width=.45\linewidth,height=5cm]{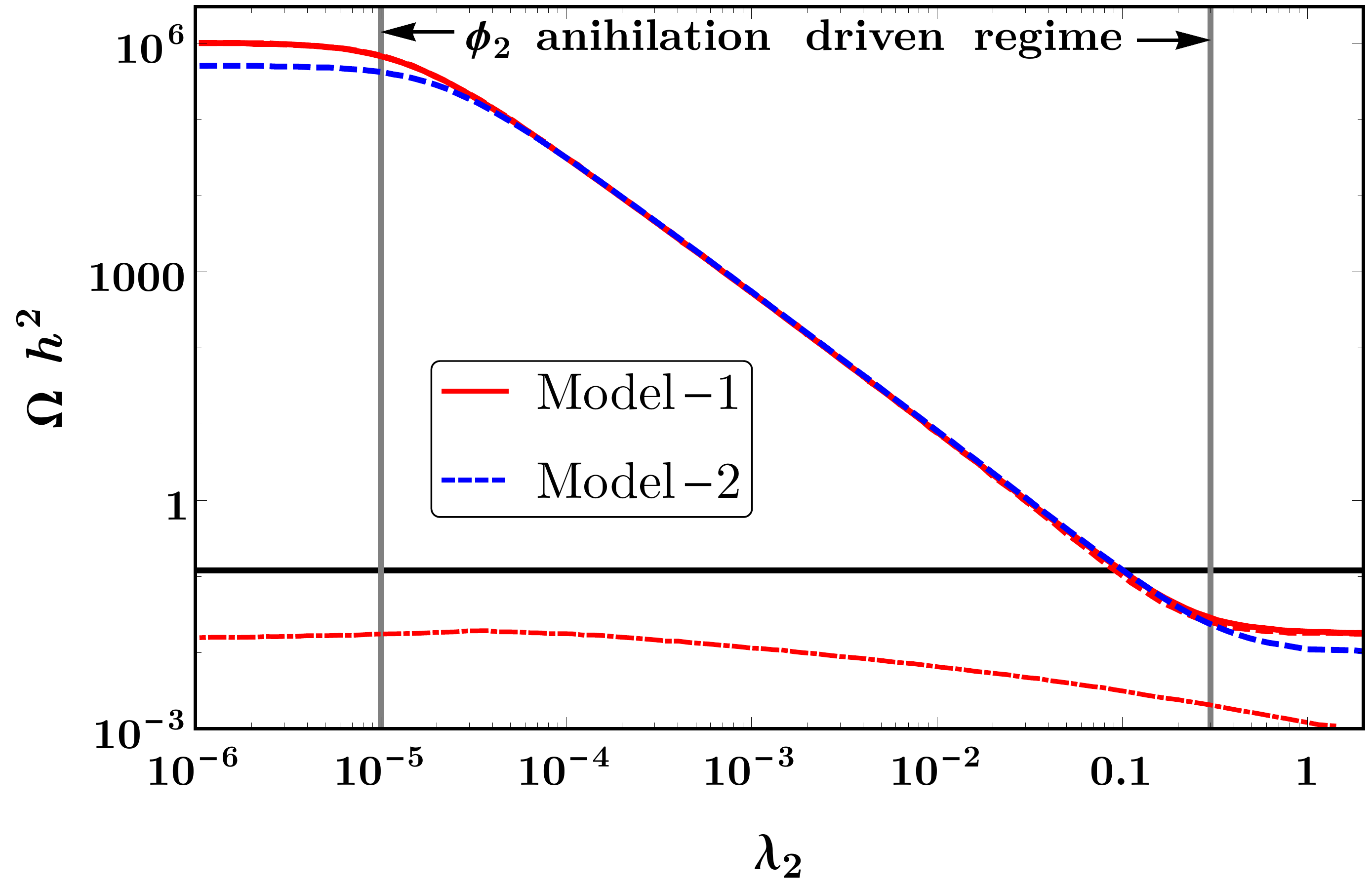}}
\vspace{-0.2cm}
\caption{
\label{fig:regime} Different phases of freeze out for the models. For model 1, in both the left and right panel red solid, dashed, and dot dashed lines represent total relic density ($\Omega_t h^2$), the contribution of $\phi_1$ ($\Omega_1 h^2$) and the contribution of $\phi_2$ ($\Omega_2 h^2$) respectively while the blue dashed lines correspond to model 2. We assume $\lambda_1 =10^{-5},~ m_1=100$ GeV, and $\delta=1\%$. The black solid band represents the central value of DM relic density, measured by Planck \cite{Aghanim:2018eyx}. (a) Exchange driven freeze out regime in $\Omega h^2  - \lambda_e$ plane. The chosen value of $\lambda_2$ is $0.5$. (b) $\phi_2$ annihilation dominated freeze out regime in $\Omega h^2  - \lambda_e$ plane. The chosen value of $\lambda_e$ is $0.5$.
}
\end{figure*}

The standard  two component DM framework with $\mathbb{Z}_2 \times \mathbb{Z}_2^{\prime} $ symmetry  is mostly disfavoured by the results of direct detection experiments except in the tuned Higgs resonance  region or for DM mass  $\gtrsim 800$ GeV \cite{Bhattacharya:2016ysw}.  Interestingly, in the regime where the annihilation cross section of  the lighter species, say $\phi_1$, is quite small  the freeze out of $\phi_1$ can still be achieved by the combined effect of  conversion of $\phi_1$ to $\phi_2$ and subsequent annihilation of $\phi_2$ to SM states.  In this region where  $\lambda _1 \ll \lambda_2, \lambda_e $ the model demonstrates the phenomenon of exchange driven freeze out explained above. Depending on  relative strength and hence the freeze out chronology, either the  exchange process or  the $\phi_2$ annihilation will set the relic density of $\phi_1$.

Assuming that  $\phi_1$ is the major constituent of DM, a detailed numerical solution is obtained using the full set of coupled Boltzmann equations involving the number densities of both $\phi_1$ and $\phi_2$. The temperature variation of relativistic degrees of freedom have been included following  \cite{Drees:2015exa}. The cross section of the all the relevant channels can be written as
\begin{align}
\label{eq:sigmav}
&\sigma v_{\phi_i \phi_i \to f \overline f}=\frac{1}{4\pi s \sqrt s} \frac{N_c{\lambda_{i}}^2 m_f^2}{(s-m_h^2)^2}(s-4m_f^2)^\frac{3}{2}
\nonumber\\
&\sigma v_{\phi_i \phi_i \rightarrow h h}=\frac{{\lambda_{i}}^2 \sqrt{s-4 m_h^2}}{16\pi s \sqrt{s}}\left(1+\frac{3m_h^2}{(s-m_h^2)}-\frac{4{\lambda_{i}} v^2}{(s-2m_h^2)}\right)^2 \nonumber \\ 
&\sigma v_{\phi_2 \phi_2 \rightarrow \phi_1 \phi_1}=\frac{\sqrt{s-4m_{\phi_1}^2}}{8\pi s \sqrt s}\left(\frac{ v^2 \lambda_{1}\lambda_{2}}{(s-m_h^2)}+\lambda_e\right)^2  \\
&\sigma v_{\phi_i \phi_i \rightarrow VV}=\frac{s_f  \sqrt{s}{\lambda_{i}}^2 \sqrt{1-4m_V^2}}{16\pi (s-m_h^2)^2} \left(1+\frac{12m_{V}^4}{s^2} -\frac{4m_{V}^2}{s}\right)  \nonumber
\end{align}
 where $ s_f =\{2,1\}~ {\rm for}~ \{W,Z\}.$
We have utilized equation \eqref{eq:sigmav} in the non-relativistic limit to solve the Boltzmann equation.  Throughout the discussion the annihilation cross section of $\phi_1$ has been kept small by fixing it at $\lambda_1 = 10^{-5}$. 

In figure \ref{fig:yt} the evolution of co-moving number densities of $\phi_1$ and $\phi_2$ is shown by red and blue lines, respectively. The dot-dashed lines represent their respective equilibrium values. For the solid curve exchange processes control the DM relic abundance. Whereas the simultaneous departure of both dashed blue and red lines from equilibrium curve indicates that the annihilation of $\phi_2$ drives the freeze out of $\phi_1$. The apparent  decrement of the $\phi_2$ number density after it departs from equilibrium curve is due to the process $\phi_2 \phi_2 \to \phi_1  \phi_1$. 
 
The two distinct possibilities of our numerical solution have been depicted in figure \ref{fig:regime}. In the left and right panel, we show variation of relic abundance with $\lambda_e$ and $\lambda_2$, respectively. In each plot, there are three distinctive regions, the region in between two gray vertical lines represents the part of the parameter space where the relic density of $\phi_1$ and hence. that of DM is set by exchange  (figure \ref{sf:Ex-regime}) or $\phi_2$ annihilation  (figure \ref{sf:2ndAnn-regime}) processes. When either of $\lambda_e$ or $\lambda_2$ become comparable to $\lambda_1$ then annihilation of $\phi_1$ regulates the relic density. In both of the panels flatness of the relic density towards the left edge before the  vertical lines implies this.   While for larger values of the $\lambda_e$ or $\lambda_2$, depending on their relative strength either exchange or $\phi_2$ annihilation will control the dark matter relic density. In figure \ref{sf:Ex-regime} for $\lambda_e \gtrsim 1$, the freeze out of $\phi_2$ control the relic density while in figure \ref{sf:2ndAnn-regime}  for $\lambda_2 \gtrsim 1$, the exchange process will drive the freeze out.

\begin{figure}[t]
\begin{center}
\includegraphics[scale=0.27]{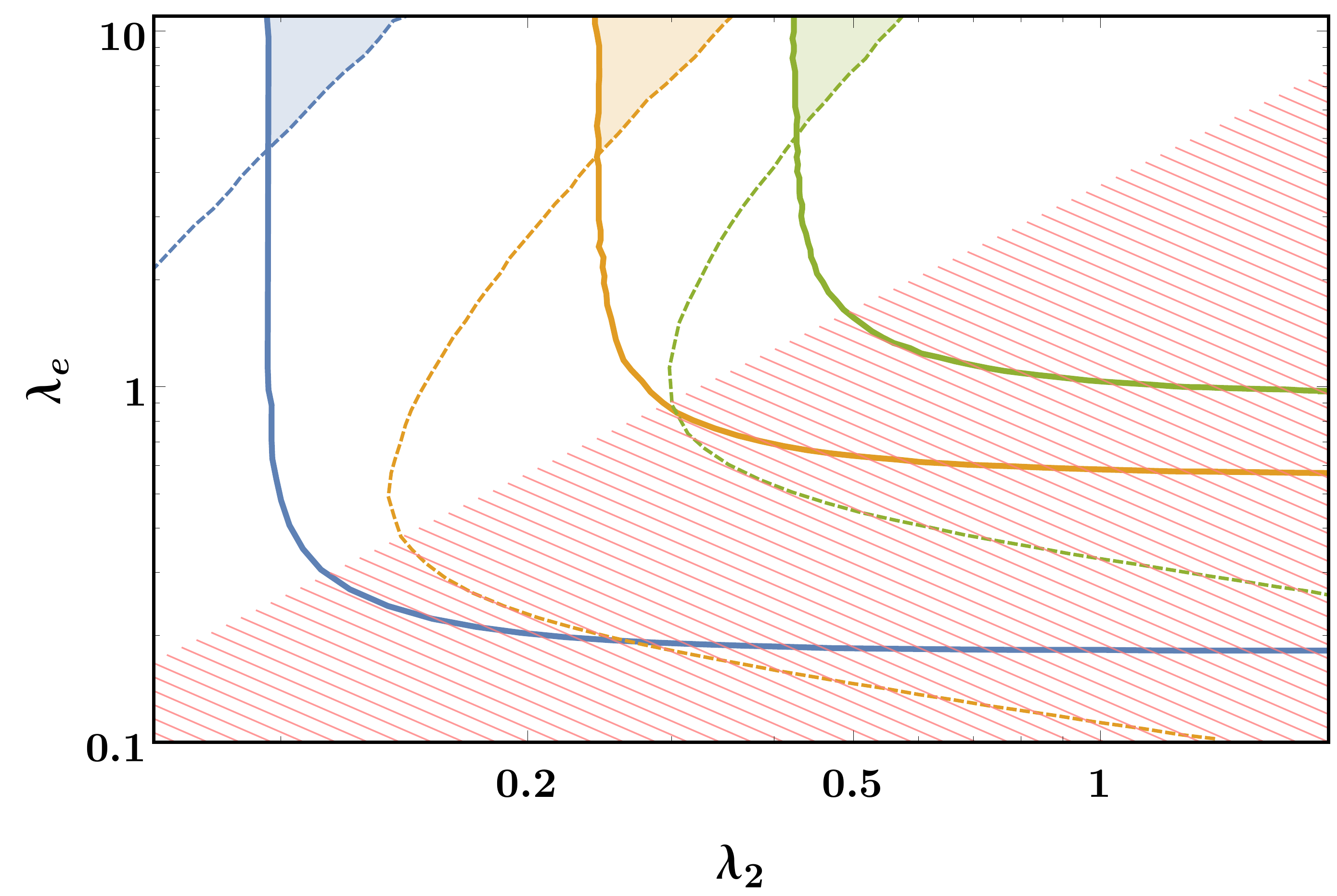}
\caption{Phenomenological consequences of model 1 in $\lambda_2-\lambda_e$ plane for $m_{1}=100,~ 300, ~\rm and~ 500$ GeV are shown by light blue, orange, and green lines respectively. The solid lines represent $\Omega_t h^2 =0.12$ contours \cite{Aghanim:2018eyx}. The dashed lines illustrate upper limit on the couplings for the aforementioned masses from XENON1T experiments. The three shaded regions show the allowed parameter space from both the over-closure of relic density and direct detection. For the hatched region, the exchange process drives the freeze out of $\phi_1$, whereas for the non-hatched region, annihilation of $\phi_2$ determines the same. The numerical value of the other relevant parameters is mentioned in figure \ref{fig:regime} .}
\label{fig:relic-z2z2p}
\end{center}
\end{figure}

In the exchange driven freeze out regime the large value of $\lambda_2$  results  in a large spin-independent $\phi_2-$nucleon cross section \cite{Goodman:1984dc, Alarcon:2012nr, Alarcon:2011zs}. This leads to  a strong constraint from direct detection limits in-spite of the fact that the $\phi_2$ is a minor constituent of the total DM relic density.  For instance, the limit on such Higgs portal coupling  from the recent results of the XENON1T experiment \cite{Aprile:2018dbl} effectively rules out this possibility in the $\sim 100$ GeV mass scale within the minimal model. However, note that the effective spin-independent $\phi_2-$nucleon cross section is given by 
\begin{equation}
\label{eq:phi2dd}
\sigma_{2,n}^{\rm eff} = \frac{\Omega_2}{\Omega_t} \sigma_{2,n}
\end{equation}
The spin independent $\phi_2-$nucleon direct detection cross section, $\sigma_{2,n}$ reads
  \begin{eqnarray} \label{eq:dd-sigma}
   \sigma_{2,n}=\frac{\lambda_{2}^2 f_n^2}{4\pi } \frac{\mu_n^2 m_n^2}{m_h^4m_2^2}~,
    \end{eqnarray}
 where $\mu_n= m_{n} m_2/(m_{n}+m_2)$,  $m_n$ is the mass of the nucleon and nucleon form factor, $f_n \approx 0.28$ \cite{Alarcon:2011zs,Alarcon:2012nr}. The first factor of the RHS of the equation \eqref{eq:phi2dd} represents the fraction of relic density shared by $\phi_2$ and as mentioned earlier $\phi_2$ comprises a very small fraction of the total DM density. Thus, in the $\phi_2$ annihilation dominated  region  the lower rate of $\phi_2$ annihilation  cross-section together with   the  reduction  factor $\Omega_2/\Omega_t \ll 1$  is enough to suppress the  effective direct detection coupling to an extent to be able to evade the present XENON1T bound. This has been demonstrated in figure \ref{fig:relic-z2z2p} where the light blue, orange, and green lines correspond to DM relic density  contours for $m_{1}=~100,~300,~\rm and ~500$ GeV, respectively. The dashed lines of figure \ref{fig:relic-z2z2p} show direct detection upper limit on $\lambda_2$ for the aforementioned masses. The  three shaded regions are consistent with both the direct detection and relic density over closure  bounds. Further, in figure \ref{fig:relic-z2z2p}, the pink hatched patch denotes the exchange driven freeze out region which remains forbidden by the direct detection bound.  Interestingly, the allowed $\phi_2$ annihilation region resurrects  the  two component  $\mathbb{Z}_2 \times \mathbb{Z}_2^{\prime}$ model with  DM masses in the $100$ GeV  range,  which is otherwise ruled out in conventional scenarios \cite{Bhattacharya:2016ysw}. It is to be noted that  in the above discussion we have chosen $\delta=1 \%$  signifying the amount of fine tuning required in this framework. 
 
In passing we note that  one can consider  the heavier of the two species as the major DM constituent which would ease the exponential suppression in equation \eqref{eq:x-section} \cite{Belanger:2011ww}. However, such a framework seems to be  mostly ruled out from direct detection constraints.

\subsection{Model 2: Single-component DM}
\label{subsec:model2}

In a setup where the two SM singlet scalars  $\phi_1$ and $\phi_2$  are charged under the same $\mathbb{Z}_2$ symmetry, keeping the rest of the SM even leads to a multipartite dark sector with a single stable DM candidate \cite{Casas:2017jjg, Bhattacharya:2017fid}. The most general renormalizable Lagrangian consistent with the charge assignment is given by
 \begin{eqnarray}
\label{eq:z22L}
-\mathcal{L}_{\rm int} &=& \Big(\frac{\lambda_{1}}{2}\phi_1^2 + \frac{\lambda_{2}}{2} \, \phi_2^2 +\lambda_{12}\, \phi_1\phi_2  \Big) H^{\dagger}H+\frac{\lambda_{1_s}}{4!}\phi_1^4 \nonumber \\ 
&+&\frac{\lambda_{2_s}}{4!}\phi_2^4+\frac{\lambda_{e_1}}{4}\phi_1^2\phi_2^2+\frac{\lambda_{e_2}}{3!}\phi_1^3\phi_2+\frac{\lambda_{e_3}}{3!}\phi_1\phi_2^3 ,   
\end{eqnarray} 
where $H$ denotes the SM Higgs doublet. The lighter state represents the stable DM candidate. In our discussion, we will assume $m_{1} < m_{2}$; therefore, $\phi_1$ represents the lightest stable particle (LSP) and is the sole DM candidate within this setup, while $\phi_2$ is the next to lightest stable particle (NLSP) that can decay to the LSP. In addition to the processes shown in figure \ref{fig:block-diagram}, co-annihilation and decay of the NLSP are the major features that distinguish this from model in section \ref{subsec:model2}. The thermally averaged cross sections for all the relevant annihilation and co-annihilation channels can be read out from \eqref{eq:sigmav}. Conventionally, the annihilation and co-annihilation set the DM relic density. However, here we will assume that these processes are suppressed enough to induce a considerable time lapse between the freeze out of the LSP annihilation process with the exchange and NLSP annihilation. This enables the  exchange processes and NLSP $(\phi_2)$ annihilation to drive the freeze out of LSP $(\phi_1).$ Furthermore to keep the discussion tractable, we assume $\lambda_{e_i} = \lambda_e$. Unlike the multi-component DM model there are two more exchange channels ($\phi_1 \, \phi_1 \to \phi_1 \, \phi_2$ and $\phi_1 \, \phi_2 \to \phi_2 \, \phi_2$) that are operative here. As is evident from equation \eqref{eq:x-section}, the exponential suppression associated with these channels would be smaller than the $\phi_1 \, \phi_1 \to \phi_2 \, \phi_2$ channel. We have incorporated all theses channels in our numerical simulation.

\begin{figure}[t]
\begin{center}
\includegraphics[scale=0.27]{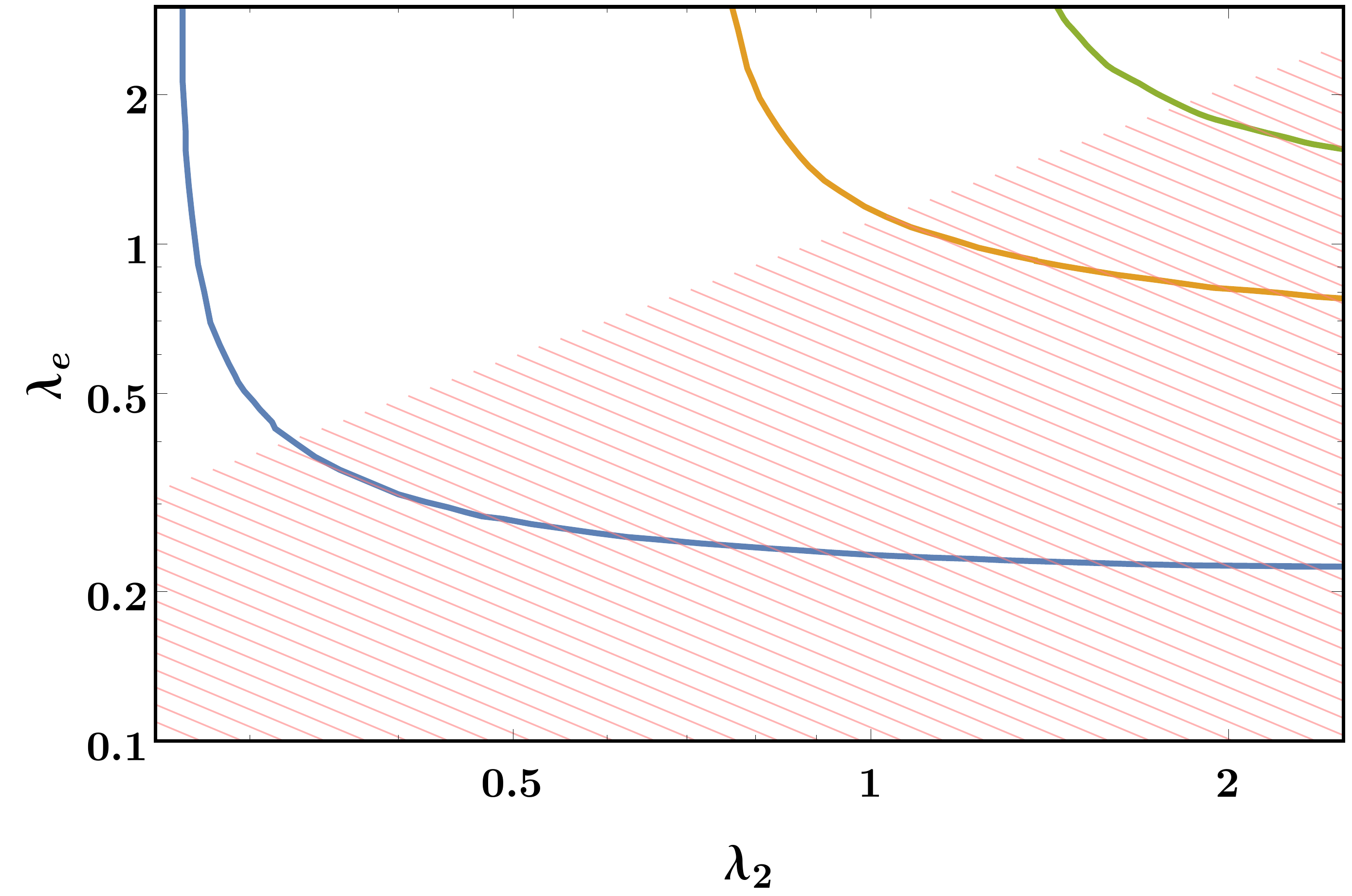}
\caption{$\Omega h^2 =0.12$ contours for model 2 in $\lambda_2-\lambda_e$ plane for $m_{1}=100,~ 300, ~\rm and ~500$ GeV are depicted by light blue, orange, and green lines, respectively. For the hatched region exchange processes drive the DM freeze out, whereas for the non-hatched region, annihilation of $\phi_2$ determines the same. We have taken $\lambda_1 =\lambda_{12}=10^{-5}$ and $\delta=5\%$.}
\label{fig:relic-z22}
\end{center}
\end{figure}

The different phases of the freeze out  shown by the blue dashed lines of figure \ref{fig:regime} which are  obtained using a full numerical simulation of the dark sector, is represented by the Lagrangian given in equation \eqref{eq:z22L}. Clearly, when $\lambda_1,\lambda_{12}<\lambda_{e/2}\sim\lambda_{2/e}$ then either exchange processes or the annihilation of $\phi_2$  predominantly  drive the thermal freeze out of the DM, hence,  we set $\lambda_1 = \lambda_{12} =10^{-5}$. The relic density allowed contours for DM masses $100,~300,~ \rm and~500$ GeV have been depicted in figure \ref{fig:relic-z22}. The hatched region denotes the exchange driven freeze out parameter space while for the non-hatched region annihilation of $\phi_2$ drives the freeze out.

The spin-independent direct detection cross section of the DM would be considerably smaller due to the smallness of the Higgs portal coupling $\lambda_1$. However, in both the exchange and $\phi_2$ annihilation driven regime the lifetime of $\phi_2$ can be quite large,  having potential implications for early Universe cosmology. For the DM mass in the 100 GeV scale  and $\delta=5\%,~\rm$  a lower limit on $\lambda_{12} \geq10^{-6} (\ll \lambda_e, \lambda_2)$  is obtained   from BBN constraints.

\section{Conclusion }
\label{sec:con}
In this paper, we propose a new paradigm of thermal freeze out of DM. In a relatively degenerate dark sector, if the annihilation cross section of the major constituent of DM  relic density is suppressed then the exchange process within the dark sector and subsequent annihilation of other species drives the freeze out. Depending on the time ordering of freeze out, either exchange processes or annihilation of second lightest species can set the relic density of DM. We have demonstrated that for simple multiparticle Higgs portal DM model, this  novel mechanism can be readily realized to saturate the DM relic density constraints while being consistent with direct detection bounds.

\acknowledgments
TNM would like to thank MHRD, Government of India for a research fellowship. TSR is partially supported by the Department of Science and Technology, Government of India, under the Grant Agreement No. IFA13-PH-74 (INSPIRE Faculty Award). TSR is grateful to the Mainz Institute of Theoretical Physics (MITP), which is a part of DFG Cluster of Excellence PRISMA$^+$ (Project ID $39083149$), for its hospitality and its partial support during the initial stage of this work.

\bibliographystyle{JHEP}
\bibliography{exchange-ref.bib}

\end{document}